# Understanding and predicting synthetic lethal genetic interactions in *Saccharomyces cerevisiae* using domain genetic interactions

Bo Li[1], Weiguo Cao[2], Jizhong Zhou[3], Feng Luo[1,*]

[1]School of Computing, Clemson University, Clemson, SC 29634, USA

[2]Department of Genetics and Biochemistry, South Carolina Experiment Station, Clemson University, Clemson, SC 29634, USA

[3]Institue for Environmental Genomics, University of Oklahoma, Norman, OK, USA

* Corresponding author.  Tel: +1 864 656 4793; Email: luofeng@clemson.edu

**Abstract**

Genetic interactions have been widely used to define functional relationships between proteins and pathways. In this study, we demonstrated that yeast synthetic lethal genetic interactions can be explained by the genetic interactions between domains of those proteins. The domain genetic interactions rarely overlap with the domain physical interactions from iPfam database and provide a complementary view about domain relationships. Moreover, we found that domains in multidomain yeast proteins contribute to their genetic interactions differently. The domain genetic interactions help more precisely define the function related to the synthetic lethal genetic interactions, and then help understand how domains contribute to different functionalities of multidomain proteins. Using the probabilities of domain genetic interactions, we were able to predict novel yeast synthetic lethal genetic interactions. Furthermore, we had also identified novel compensatory pathways from the predicted synthetic lethal genetic interactions. Our study significantly improved the understanding of yeast mulitdomain proteins, the synthetic lethal genetic interactions and the functional relationships between proteins and pathways.

# Introduction

Defining the functional relationships between proteins is one of the important tasks in the post-genomic era. A classical approach to understand gene functional relationships is producing combination mutant in two genes to observe genetic interactions [1]. Genetic interaction refers to the phenomenon in which the combined effect of mutations of two genes differs from individual effects of each mutation [2]. In the extreme cases, mutation of two nonessential genes could lead to a lethal phenotype. This kind of genetic interactions is referred as synthetic lethal genetic interactions (SLGIs). The genome-wide SLGIs have attracted much attention as they are capable of defining the genome-wide functional relationships between proteins, pathways and complexes [2-4]. The SLGIs also have potential for finding drug target or drug combinations [5].

Representing the structures and functions of proteins, protein domains are usually regarded as building blocks of proteins and are conserved during evolution. The mutation of a gene causes the loss of function of its protein product, which may accredit to the loss of protein domains in the protein product. Then, the effect of the mutation of two genes may be caused by the loss of protein domain combinations in both protein products. We refer the phenomenon in which combined effect of mutations of two domains in two proteins differs from individual effects of mutation of each domain as domain genetic interactions. The domain genetic interactions may correlate to SLGIs. We may use the domain genetic interactions to explain and predict the SLGIs. Furthermore, in multidomain proteins, different domains may fulfill different functions independently or collaboratively. Although genetic interaction analysis provides a promising method to understand the functional relationship between proteins [6], we cannot tell the contributions of different domains to certain functionality of multidomain proteins based on their genetic interactions. Studying the domain genetic interactions may help elucidate the SLGIs between multidomain proteins at domain level.

Genetic interactions are usually identified by mutant screens [7]. Recently, high throughput technologies, such as synthetic genetic arrays (SGA) [8] or synthetic lethal analysis by microarrays (SLAM) [9], have been developed for parallel and massive detection of genetic interactions. However, even with high throughput methods, experimental discovery of SLGIs is still overwhelming. Therefore, it is of interest to computationally predict SLGIs. Several computational approaches have been proposed for the prediction of SLGIs [10-15] and various features, such as protein interactions, gene expression, functional annotation, gene location, protein network characteristics, and genetic phenotype, had been utilized by these methods. Recently, Park et al. [15] also used protein domains as one of the features to predict interactions, including genetic interactions.

In our previous study [16], we performed a cross validation study on predicting SLGIs using protein domains as features. Our support vector machine (SVM) classifiers were able to achieve high performance with AUC (The area under the ROC curve) as 0.9272. These results suggested that using domain information may catch the genetic relationships between proteins. In this study, we first applied the Maximum Likelihood estimation (MLE) approach to estimate the probabilities of domain genetic interactions from yeast SLGIs. The EM algorithm developed by Dempster et al. [17] has been used to iteratively compute the maximum likelihood. Recently, the MLE method had been used to infer domain physical interactions from protein physical interactions [18, 19]. We identified significant domain genetic interactions, which rarely overlap with the domain physical interactions from iPfam database [20]. We also showed that domains in multidomain yeast proteins contribute to their genetic interactions differently. Then, we used the probabilistic model to predict yeast SLGIs using the probabilities of domain genetic interactions. We were able to predict novel yeast SLGIs using probabilities of domain genetic interactions, which demonstrate the ability of protein domains in predicting SLGIs. We were also able to apply our predicted yeast SLGIs to understand the compensatory pathways. A complete description of our results and methods is given in the sections below.

## Results

**Identification of Significant Domain Genetic Interactions Using Maximum Likelihood Estimation**

By assuming the independences among domain genetic interactions, we estimated the probabilities of domain genetic interactions using the Maximum Likelihood estimation (MLE) method. The probability of a domain pair indicates its propensity to genetically interact. The high probability of a domain pair imply that number of SLGIs including the domain interactions is high comparing to the number of protein pairs including the domain interaction that are not genetically interacted. Then, we calculated an evidence score E [19] for each domain genetic interaction to estimate its relative significance. The E scores are not the standard likelihood ratio test as they are calculated using only the positive data [19]. The E scores indicate the relative significances of the domain pairs in the SLGIs. The probability and the E score give different view about the significances of domain genetic interactions. The high probabilities do not always correlate to high E score. For example, there are a large number of protein pairs including a single domain pair. If very few of them are SLGIs, the probability of the domain pair will be low. However, if that domain pair is the only domain pair in those SLGIs, it will have high E score. On the other hand, in case there are a large number of protein pairs containing the same multiple domain pairs, if most of the protein pairs are SLGIs, the probabilities of those domain pairs will be high. However, the E scores of those domain pairs will be relatively low as removing any of those domain pairs will have limited effect on the probabilities of SLGIs. Therefore, we used both the probability and E score to select significant domain genetic interactions.

We first applied the MLE approach to the 7475 genetic interactions with both proteins having protein domains. We were able to assign probabilities and E scores to 11,189 domain pairs. Those domain genetic interactions are available on our website, www.genenetworks.net, for searching and downloading. We first selected the significant domain genetic interactions with E scores greater than or equal to 2.0, which corresponds to an approximate seven fold drop of the probability of SLGIs if this domain genetic interactions is excluded. Then, we selected the domain genetic interactions with probability greater than 0.5 even though they have low E score values. Totally, we obtained 3848 domain genetic interactions of 1027 domains. Table II lists top ten domain genetic interactions with the highest E scores. The probabilities of those ten domain genetic interactions vary from 0.013 to 0.8. The prefoldin domain dominates in the top ten domain genetic interactions (7 of 10). Our results are consistent with the findings of Le Meur and Gentleman [21], in which they showed that the prefoldin complex is in 9 of their top 10 pairs synthetic multi-protein complexes genetic interactions.

**Domain Genetic Interactions Rarely Overlap Domain Physical Interactions**

To investigate the relationship between domain genetic interactions and the domain physical interactions, we compared the 3848 significant domain genetic interactions with the domain physical interactions from the iPfam database [20]. There are 4030 domain physical interactions of 1867 domains in iPfam database (2008 version). Among them, 1556 domain physical interactions of 1048 domains exist in yeast proteins we studied. There are 422 domains in both the genetic and physical interactions. We applied the Fisher's exact test in R to examine if domains participating genetic interactions are independent from domains participating in physical interactions. We obtained a p-value of 0.00025, which indicated a significant difference between two sets of protein domains. Furthermore, there are only 70 domain pairs overlapped between the 1556 domain physical interactions and 3848 domain genetic interactions. The Fisher's exact test obtained a p-value of 0, which showed that the domain genetic interactions significantly differ from the domain physical interactions. Thus, the domain genetic interaction is a new type of relationship among domains and can provide a complementary view about the relationships between domains.

**The Properties of Domain Genetic Interaction Network**

To obtain an overview of the domain genetic interactions, we modeled the domain genetic interactions as a network, in which each node represents a domain and each link represents a genetic interaction between two domains. Then, we examined the properties of this domain genetic interaction network. The average connectivity of the network is 7.392. The domain PF00022 (Actin) has the highest connections of 186. The average shortest path among all nodes is 3.31. Comparing the size of nodes, this number is very small. Furthermore, the average node clustering coefficient of the network is 0.159. These properties indicated that the domain genetic interaction network has the small world property [22]. The analysis of the connectivity distribution of this network showed a power-law distribution with an exponent degree of 1.45 (Figure 1), which indicated that the domain genetic interaction network is a scale free network [23]. These results demonstrated that the domain genetic interaction network follows the common principles of biological networks [24].

**Domain Genetic Interactions in the SLGIs between Multidomain Proteins**

Most proteins are multidomain proteins, which are created as a result of different genetic events, such as insertions and duplications [25, 26]. Multidomain proteins may have different functionalities due to different domains. Our identification of domain genetic interactions helps understand the domains that contribute to functionality defined by the SLGIs, and then help elucidate the functional relationships between proteins at domain level from their genetic interactions.

Figure 2 shows examples of domain genetic interactions in three SLGIs between yeast multidomain proteins. Figure 2A shows the domain genetic interactions between SGS1 and TOP1. The SGS1 has three domains and TOP1 has two domains. Only one of the three domains of SGS1, PF00570 (HRDC domain), has high probabilities to interact with two domains of TOP1: PF01028 (Eukaryotic DNA topoisomerase I catalytic core domain) and PF02919 (Eukaryotic DNA topoisomerase I DNA binding domain). The other two domains: PF00270 (DEAD/DEAH domain) and PF00271 (Helicase conserved C-terminal domain) of SGS1 have no genetic interaction with two domains of TOP1 (very low probabilities). Figure 2B shows the domain genetic interactions between TOP3 and TOP1. Both domains of TOP3: PF01131 (DNA topoisomerase domain) and PF01751 (Toprim domain) show high probabilities of genetic interaction with two domains of TOP1. Figure 2C shows the domain genetic interactions between RAD5 and RAD50. The RAD5 has four domains and RAD50 has two domains. The genetic interaction between RAD5 and RAD50 is mainly due to a single domain genetic interaction between PF08797 (HIRAN domain) of RAD5 and PF04423 (zinc hook domain) of RAD50. The PF00097 (Zinc finger domain), PF00176 (SNF2 family N-terminal domain) and PF00271 (Helicase conserved C-terminal domain) of RAD5 and PF02463 (RecF/RecN/SMC N terminal domain) of RAD50 have low contribution to the genetic interaction between RAD5 and

RAD50. Those examples show different domain genetic interaction architectures in SLGIs between yeast multidomaon proteins. The domain genetic interactions may exist between all domains of two proteins (TOP3 and TOP1), or between part of domains of one protein and all domains of the other protein (SGS1 and TOP1), or between part of domains of one protein and part of domains of the other protein (RAD5 and RAD50). The domain genetic interactions are able to help understand functional relationships between multidomain proteins at domain level.

We also investigated the domain genetic interactions of SLGIs between SGS1 and other proteins. We found that only the PF00570 (HRDC domain) of SGS1 has significant genetic interactions with other domains. The results implied that certain functionality of SGS1 may be only due to its HRDC domain, rather than its DEAD/DEAH domain and helicase conserved C-terminal domain. Previous study showed that the HRDC domain of SGS1 is required for its cellular functions involving topoisomerases [27]. Thus, the domain genetic interaction analysis can help understand how domains contribute to the different functionalities of multidomain proteins.

**Prediction and Validation of Genome-wide SLGIs Using Protein Domains**

Having established that there is a strong correlation between domain genetic interactions and SLGIs, we explored to predict the probabilities of protein pairs to be SLGIs using probabilities of domain genetic interactions. We were able to assign 599752 protein pairs with probabilities greater than 0. Supplemental Table I lists the number of SLGIs predicted by MLE approach at different probability cutoffs. All predicted SLGIs are hosted on our website for searching and downloading.

We then compared the correlation coefficients of gene expressions of predicted SLGIs to those of known SLGIs and those of all possible protein pairs. We used a yeast cell cycle gene expression data [28], which contains 77 data points. We calculated the T-score and P-value for the null hypothesis that there is no difference between the means of predicted SLGIs and the means of known SLGIs and the null hypothesis that there is no difference between the means of predicted SLGIs and the means of all pairs. The results are shown in Supplemental Table II. The correlation coefficients of gene expression of predicted SLGIs with threshold greater than 0.3 are similar to those of known SLGIs. The gene expression correlation coefficients of predicted SLGIs using different thresholds significantly differ from those of all pairs except for SLGIs with probabilities greater than 0.85, which have only small numbers of SLGIs. Those results indicated that the correlation coefficients of gene expressions of predicted SLGIs are similar to those of known SLGIs, rather than to those of random pairs. Recently, it was reported that the SLGIs are likely to have similar GO annotations [29]. We studied the distribution of similarities of Gene Ontology (GO) annotations between predicted SLGIs and also compared them to those of known SLGIs and all possible protein pairs. As shown in Supplemental Table III-V, the mean similarities of GO annotations of predicted SLGIs always significantly differ from those of all pairs. At certain probability thresholds, the mean similarities of GO annotations of predicted SLGIs show no significant differences from those of known SLGIs. Those thresholds are 0.35 for biological process and cellular component and 0.25 for molecular function. As the probability thresholds increase, the mean similarities of GO annotations also increase, which will make them differ from those of known SLGIs. The studies of GO annotations similarities and expression correlation coefficients showed that the predicted SLGIs at probability threshold around 0.3 are similar to experimentally obtained SLGIs.

**Novel SLGIs predicted by MLE Approach**

The MLE approach was able to predict novel SLGIs. Table II lists 17 novel SLGIs (not included in our training data) with probability >0.9. We predicted the MYO4/DYN1 pair to be SLGI with the highest probability of 0.9895. The MYO4 is one of two type V myosin motors. The other one is MYO2, which is known to genetically interact with DYN1 [29]. Among 17 SLGIs, 12 SLGIs are between transcription initiation factor genes and genes from RNA polymerase complex. Previously, many SLGIs between transcription initiation factor genes and RNA polymerase genes have already been reported [30-32]. We expected our novel SLGIs to help further elucidate the transcription machine. We then investigated genes involved in cellular response to stresses caused by DNA damage. We downloaded a list of 116 DNA repair and recombination genes from Kyoto Encyclopedia of Genes and Genomes (KEGG) database [33]. Then, we extracted SLGIs in which at least one protein of SLGI pair are related to DNA repair. Figure 3 shows the DNA repair related SLGIs with probability>0.7. Of a total of 133 SLGIs, 22 SLGIs are novel. Some of new predicted SLGIs are supported by previous studies. For example, the TOP3 and RAD1 double mutant has shown extreme synergistic growth defects in a previous study [34]. A recent study showed that the RTT109 and YKu70 double mutant exhibits synergistic defects under hydroxyurea treatment [35]. The PAP2 and POL2 were also shown to genetically interact at high temperatures [36].

**Compensatory Pathways from Predicted SLGIs**

Protein pathways are a part of gene network in the cell that can accomplish certain functionality. The SLGIs have been proposed to have high probability of occurrence in compensatory pathways [37]. Thus, the SLGIs within the pathways are rare and the SLGIs between pathways are significantly abundant. Identification compensatory pathways from synthetic lethal genetic interactions can be a powerful way to understanding cellular functional relationships. We expected our new predictions will increase the ability of understanding compensatory pathways. We applied the algorithm of Ma [38] to identify compensatory pathways from 7583 predicted SLGIs with probability higher than 0.3. Among 7583 SLGIs, 4497 SLGIs are novel predictions. Although Ma et al. have shown that physical interactions are enriched in discovered pathways [38], there is no assumption that proteins in those pathways are physically interacting. Totally, we obtained 167 pairs of compensatory pathways, which include 638 proteins and 3535 SLGIs. Then, we examined the GO term co-occurrences in each pathway using the SGD GO Term Finder [39]. The GO Term Finder calculated the p-values that reflect the probability of observing the co-occurrence of proteins with a given GO term in a certain pathway by chance based on a binomial distribution. Among 167 compensatory pathway pairs, 153 pairs are found significantly enriched GO terms with p-value less than 0.05. All 167 pairs of compensatory pathways and their most significant GO terms are listed on our www.genenetworks.net website.

Figure 4A and 4B lists two pairs of compensatory pathways related to DNA double strand break (DSB) repair. Among total 28 SLGIs between those two compensatory pathways, only 12 of them are known. The DSB is a kind of lethal DNA damage in which both strands of double helix are cleaved. The cell maintains multiple mechanisms to repair double strand breaks with the homologous recombination (HR) and non-homologous end joining (NHEJ) as two major mechanisms. Pathways involving those two mechanisms had shown to be compensatory to each other in Drosophila [40]. The pathways on the left of two compensatory pairs involve in NHEJ. The KU complex (YKU80 and YKU70) is the damage detector of NHEJ [41]. Mutation of nucleoporins NUP84 and NUP133 was reported to be hypersensitive to DNA damage [42]. Furthermore, the nucleoporins (NUP84 and NUP133) had been reported to colocalize and coimmunoprecipitate with Slx5/Slx8 [43], which regulate a DNA repair pathway that counteract

Rad5-independent HR [44]. Our discovery suggested that the nucleoporins (NUP84 and NUP133) may regulate the NHEJ pathway through Ku complex for double strand breaks repair. The RAD24 is a DNA damage checkpoint protein [45]. Studies also had shown that the YKU80 and RAD24 are in the same NHEJ pathways [45] to repair irradiation and methylmethanesulphonate (MMS) damages. The CSM3 is a DNA replication checkpoint protein. Genetic interaction between RAD24 and CSM3 may indicate the pathway on the right side of Figure 5 A can actually become two pathways: one is YKU80, RAD24, NUP84; the other is YKU80, CSM3 and NUP84.

The pathways on the right of those two compensatory pairs of Figure 4 A and 4B involve in HR. On the right of Figure 4A, the pathway involves three proteins: RAD57, RAD51 and DMC1. The DMC1 and RAD51 are known to form a complex [46] and have roles in recombination [47]. RAD51 and RAD57 are in the same protein family and it has been shown that RAD57/RAD55 bind with RAD51 [48]. On the right of Figure 4B, the pathway involves four proteins: SGS1, RRP6, MRE11 and RAD52. The MRE11 and SGS1 are part of a two-step mechanism to initial HR [49]. The RAD52 plays a major role in the single strand annealing and strand exchange.

Figure 4C also shows two compensatory DNA repair pathways. Among 18 SLGIs between two pathways, 11 are novel predictions. Many previous studies have supported our prediction of compensatory functionalities of those two pathways. Guillet and Boiteus [50] reported that the APN2 and MUS81/MMS4 have overlapping function to repair 3'-blocked single strand breaks (SSBs). Vance and Wilson [51] showed that TDP1and RAD1 function as redundant pathways. SGS1/TOP3 had also been showed to overlap functionally with the MMS4/MUS81 [52]. Proteins in these compensatory pathways involve in many DNA-repair pathways, such as base excision repair (BER), nucleotide excision repair (NER) and HR. Studies already showed that BER, NER and HR pathways have overlapping specificities [53]. Our prediction of these compensatory pathways may help to understand the overlapping functionalities among BER, NER and HR pathways.

Another interesting compensatory pathway pair is related to hydroperoxides response in the cell. As shown in Figure 4D, 11 of 15 SLGIs connecting two pathways are novel predictions. The three proteins (TSA1, TSA2 and PRX1) on the right are all thioredoxin peroxidase. They play a role in reduction of hydroperocxides in cytoplasm (TSA1 and TSA2) and mitochondrion (PRX1). Meanwhile, the five proteins (RAD52, RAD5, MDM31, MDM32, and MRE11) in the left pathway are related to DNA repair. RAD52, RAD5 and MRE11 are known related to double strand break repair [54-56]. The MDM31 and MDM32 relate the stability of mitochondrial DNA [57]. Those five proteins may relate to repair the damage created by the hydroperoxides. This compensatory pathway pair implies that removing hydroperoxides or repairing damage is alternative strategies for cells to survive from hydropeoxides.

## Discussion

In this study, we demonstrated that the synthetic lethal genetic interactions between two proteins can be described by the genetic interactions between domains of those proteins. We applied a probabilistic model to successfully identify significant domain genetic interactions. The significant domain genetic interactions rarely overlap with the domain physical interactions from iPfam [20], which suggests that the domain genetic interactions and domain physical interactions are complementary to each other. The domain genetic interactions offer a better understanding of the relationship among domains, and then among proteins. Furthermore, with the identified domain genetic interactions, we showed that the contributions of domains in a multidomain protein to its genetic interactions are significantly different. The domain genetic interactions will help to decipher the domains that perform the function related to the genetic interaction. For example, analyzing the domain genetic interactions confirms that only the PF00570 (HRDC domain) of SGS1 involve in the cellular functions that are compensatory to topoisomerases.

Our identification of domain genetic interactions and prediction of SLGIs is not complete. First, our training data is limited. It is believed that the available genetic interactions are just a small fraction of the whole genetic interactions [58]. Second, the training data largely come from several high-throughput genetic interaction screens, which emphasized certain biological processes, such as DNA repair. Thus, the probability of domain genetic interaction may be effected by the future available SLGIs. Moreover, we should keep in mind that our predicted SLGIs require further validation to exclude false positives. The MLE approach assumes the independence of domain genetic interactions. However, there may be dependence between domain genetic interactions. An apparent extension of this work is to consider the dependences among cooperative domains in multidomain proteins. Due to those limitations, the performance of predicting genome-wide SLGIs using protein domains may be not as good as other classifiers [11].

In summary, our study significantly improved the understanding of different domains in mulitdomain proteins. The identification of domain genetic interactions helps the understanding of originality of functional relationship in SLGIs at domain level. Furthermore, our prediction of SLGIs expanded the ability to elucidate the functional relationships between proteins and pathways.

# Methods

**Source of Data**

We collected the protein domain data from Pfam (Protein families database) [59]. The Pfam database provides two types of protein family data. Pfam-A domains are manually curated while Pfam-B domains are automatically generated. In our study, only Pfam-A domains were considered. The total number of selected Pfam-A domains for yeast is 2289. We downloaded the genetic interactions of yeast from the Saccharomyces Genome Database (SGD) [39] (July 2009 version), which were compiled from different biological experiments [8, 60-66]. Then, we extracted synthetic lethal interaction data set from the file containing all the genetic interactions. After removing replicates, there were totally 10977 pairs of synthetic lethal interactions of 2640 proteins. We removed protein pairs from study if either protein in the pair does not contain any domain. Eventually, we obtained 7475 synthetic lethal interactions among 2008 proteins. There were 1357 unique domains in those 7475 SLGIs. The information of SLGIs used in this study is summarized in Table III.

**Estimation of probabilities and significances of domain genetic interactions**

We treated the protein genetic interactions $P_{m,n}$ and domain genetic interaction $D_{i,j}$ as random variables. The $P_{m,n}=1$ if two proteins $i$ and $j$ genetically interact and $P_{m,n}=0$ otherwise. The $D_{i,j}=1$ if two domain $i$ and $j$ genetically interact and $D_{i,j}=0$ otherwise. We estimated the probabilities of potential domain interactions $\Pr(D_{i,j}=1)$ by maximizing the likelihood of observed genetic interactions using the Expectation-Maximization (EM) algorithm [17-19]. The EM algorithm iteratively estimates the maximum likelihood of the 'complete data' that combine the observed data and unobserved data. Here, the protein genetic interactions and the domain information of proteins are our observed data and the domain genetic interactions are our unobserved data. By assuming the independences among domain genetic interactions, the likelihood of observed protein genetic interactions based on domain genetic interactions can be obtained as:

$$L = \prod_{i,j} \Pr(D_{i,j} = 1)^{M_{i,j}+a}(1-\Pr(D_{i,j} = 1))^{N_{i,j}+K_{i,j}+b} \quad (1)$$

where $M_{i,j}$ be the number of genetic interacting pairs between domain i and j in all protein genetic interactions; $N_{i,j}$ be the number of non genetic interacting domain pairs between domain i and j in protein genetic interactions; and $K_{i,j}$ be the number of non genetic interacting protein pairs including domain i in one protein and j in the other one. The value of $K_{i,j}$ is counted from all possible protein pairs with domain i in one protein and j in the other one with excluding the known genetic interacting protein pairs. The $K_{i,j}$ will keep unchanged during EM computation. The *a* and *b* are pseudo counts to avoid the $\Pr(D_{i,j}=1)$ or $\Pr(D_{i,j}=0)$ to be zero when instances of domains i and j are rare. We set both *a* and *b* to 1 in our calculation. Initially, the $M_{i,j}$ was set to the number of genetic interactions between domain i and j in experimental genetic interactions; $N_{i,j}$ is set to 0. And $\Pr(D_{i,j}=1)$ was initialized as following:

$$\Pr(D_{i,j} = 1) = \frac{M_{i,j}}{M_{i,j} + N_{i,j} + K_{i,j}} \quad (2)$$

In each Expectation step of EM algorithm, we first estimated the expected values of $E[M_{i,j}]$ and $E[N_{i,j}]$ [19] using the current $\Pr(D_{i,j}=1)$:

$$E[M_{i,j}] = \sum_{m,n} E[\frac{\Pr(D_{i,j}=1)}{1-\prod_{i \subset D(m), j \subset D(n)}(1-\Pr(D_{i,j}=1))}] \quad (3)$$

$$E[M_{i,j}] = \sum_{m,n}(1-E[\frac{\Pr(D_{i,j}=1)}{1-\prod_{i \subset D(m), j \subset D(n)}(1-\Pr(D_{i,j}=1))}]) \quad (4)$$

Then, we calculate the $\Pr(D_{i,j}=1)$ using the $E[M_{i,j}]$ and $E[N_{i,j}]$ as following (Maximization step):

$$\Pr(D_{i,j}=1) = \frac{E[M_{i,j}]+a}{E[M_{i,j}]+E[N_{i,j}]+K_{i,j}+a+b} \quad (5)$$

The EM algorithm was iterated the above Expectation and Maximization steps till the change of likelihood L is less than a pre-defined small value.

The evidence score $E_{i,j}$ [19] of domain pair i and j is defined as the ratio between the probability that a pair of proteins, m and n, genetically interact given that the pair of domains, i and j, genetically interact and the probability that a pair of proteins, m and n, genetically interact given that the pair of domains, i and j, do not genetically interact:

$$E_{i,j} = \sum_{\substack{i \in D(m) \\ j \in D(n)}} \log \frac{\Pr(P_{m,n} = 1 | i, j \text{ interact})}{\Pr(P_{m,n} = 1 | i, j \text{ not interact})} = \sum_{\substack{i \in D(m) \\ j \in D(n)}} \log \frac{1 - \prod_{k \in D(m), l \in D(n)} (1 - \Pr(D_{k,l} = 1))}{1 - \prod_{k \in D(m), l \in D(n)} (1 - \overline{\Pr(D_{k,l}^{i,j} = 1)})} \quad (6)$$

The $\overline{\Pr(D_{k,l}^{i,j} = 1)}$ denotes the probability of genetic interacting between domains k and l given that the domains i and j do not genetically interact.

**Prediction of SLGIs using probabilities of domain genetic interactions**

We assumed that two proteins genetically interact ($P_{m,n}$ =1) if and only if at least one domain pair from the two proteins genetically interact ($D_{i,j}$ =1). Then, we calculated the probability of two proteins genetically interacting $\Pr(P_{m,n}$ =1) as following:

$$\Pr(P_{m,n} = 1) = 1.0 - \prod_{\substack{i \in D(m) \\ j \in D(n)}} (1 - \Pr(D_{i,j} = 1)) \quad (7)$$

A pair of proteins was predicted to be SLGI only if its probability is higher than a predefined threshold.

## Authors' Contributions

FL designed research. BL implemented the idea. FL, BL, WGC and JZH analyzed the results. All authors have read and approved the final manuscript.

## Acknowledgements

This research was supported in part by the Institute for Modeling and Simulation Applications at Clemson University and by the NSF funding 0960586 to FL.

**Figure Legends**

**Figure 1. The log-log plot of degree distribution P(K) of domain genetic interaction network.** The linear characteristics indicated by red line ($y=x^{-1.45}$) imply that P(K) follows a power low.

**Figure 2. The domain genetic interactions in SLGIs between multidomain proteins.** The thick lines indicate significant domain genetic interactions. The thin lines indicate the domain genetic interactions with low probability. The probabilities of domain genetic interactions are labeled besides the line.

**Figure 3. The SLGIs (probabilities>0.7) related to DNA repair proteins.** The cycles indicate DNA repair proteins and the triangle indicate non DNA repair proteins. Each genetic interaction involves at least one DNA repair protein. The wide links represent new genetic interactions. The figure was produced using Cytoscape [67].

**Figure 4. Compensatory pathways identified from predicted SLGIs.** (A, B) two compensatory pathways related to DNA double strand breaks repairs. Pathways on the left of A and B belong to non-homologous end joining. Pathways on the right of A and B belong to homologous recombination. (C) A compensatory pathways related to DNA repairs. (D) A compensatory pathways related to hydroperoxides response in the cell. Dashed lines indicate known SLGIs and solid lines indicated novel predicted SLGIs. The figure was produced using Cytoscape [67].

**Figure 1.**

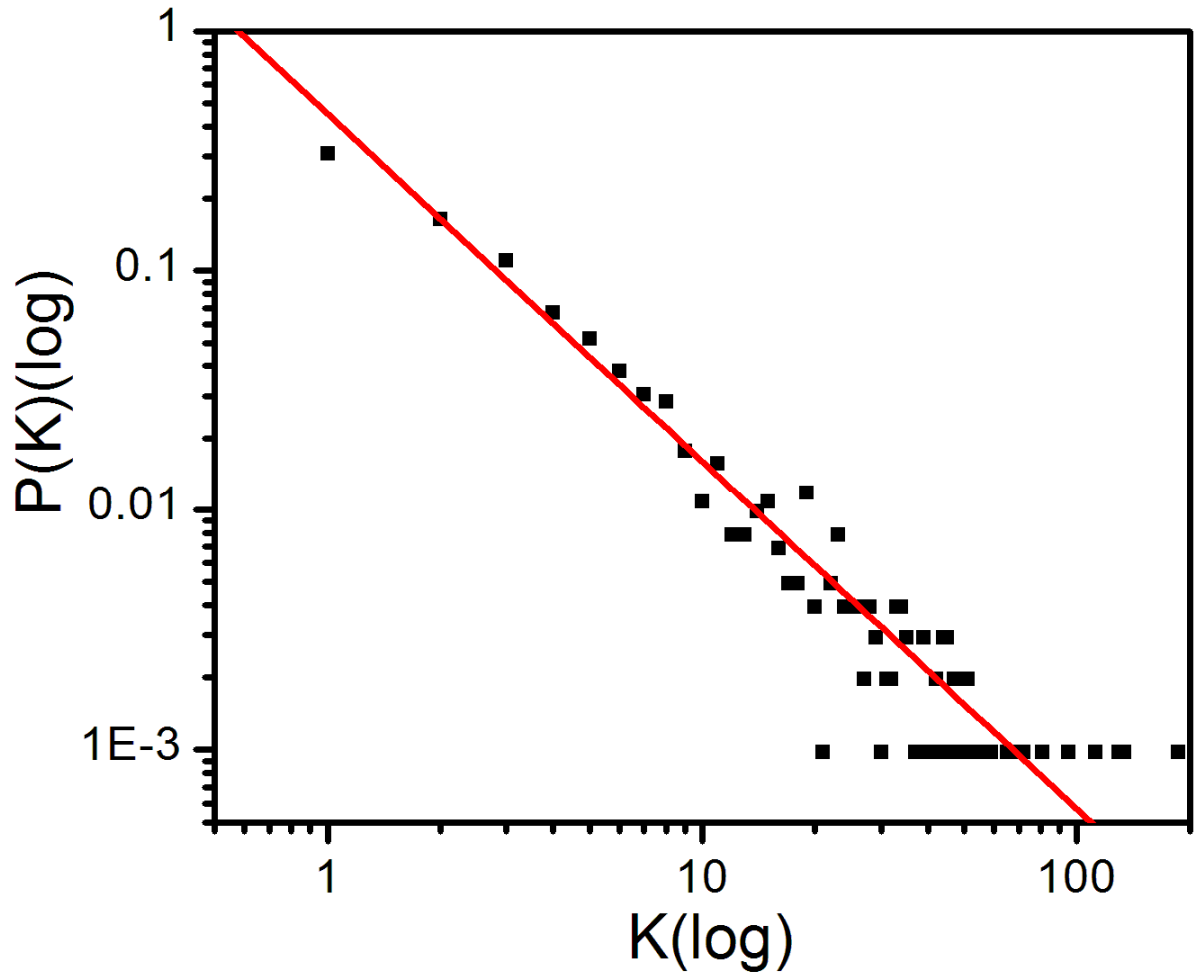

**Figure 2.**

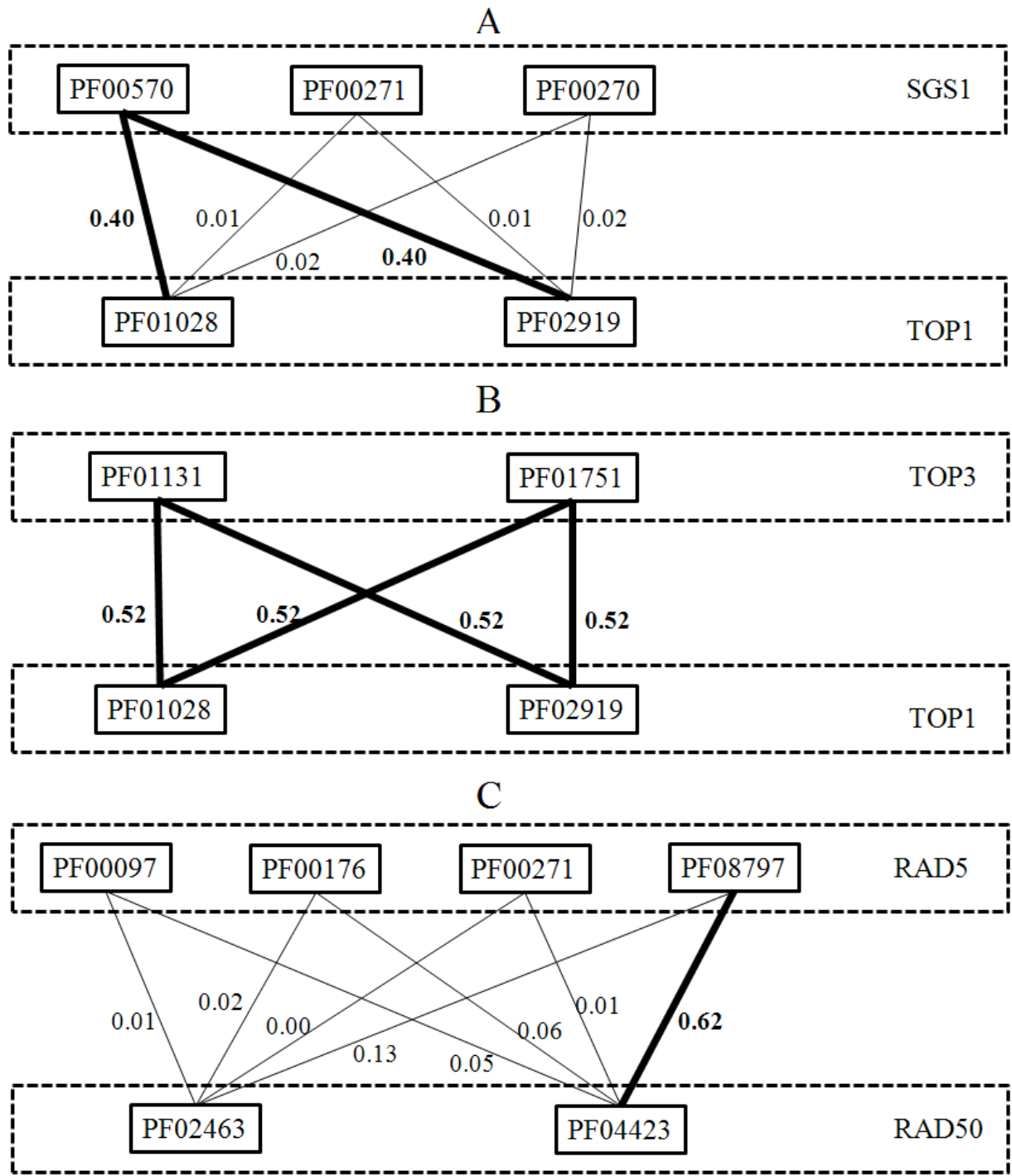

**Figure 3.**

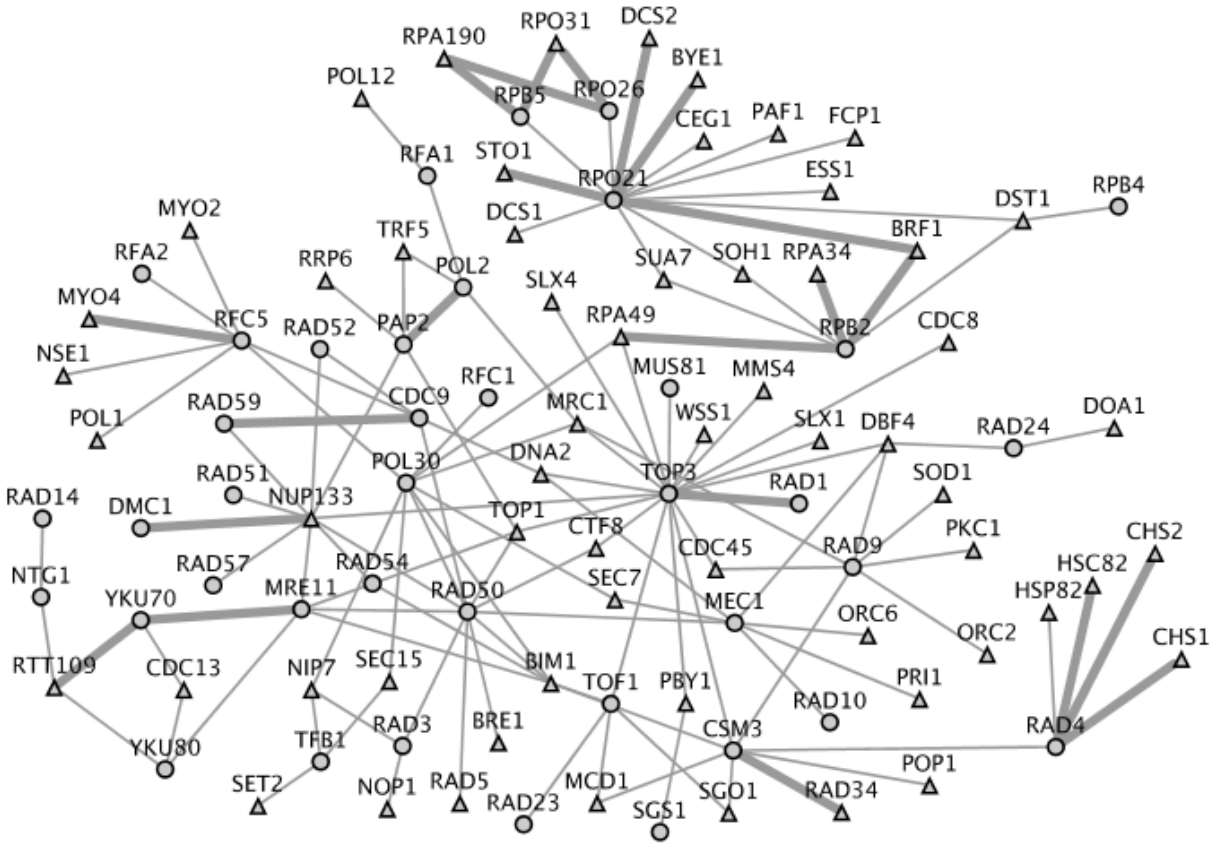

**Figure 4.**

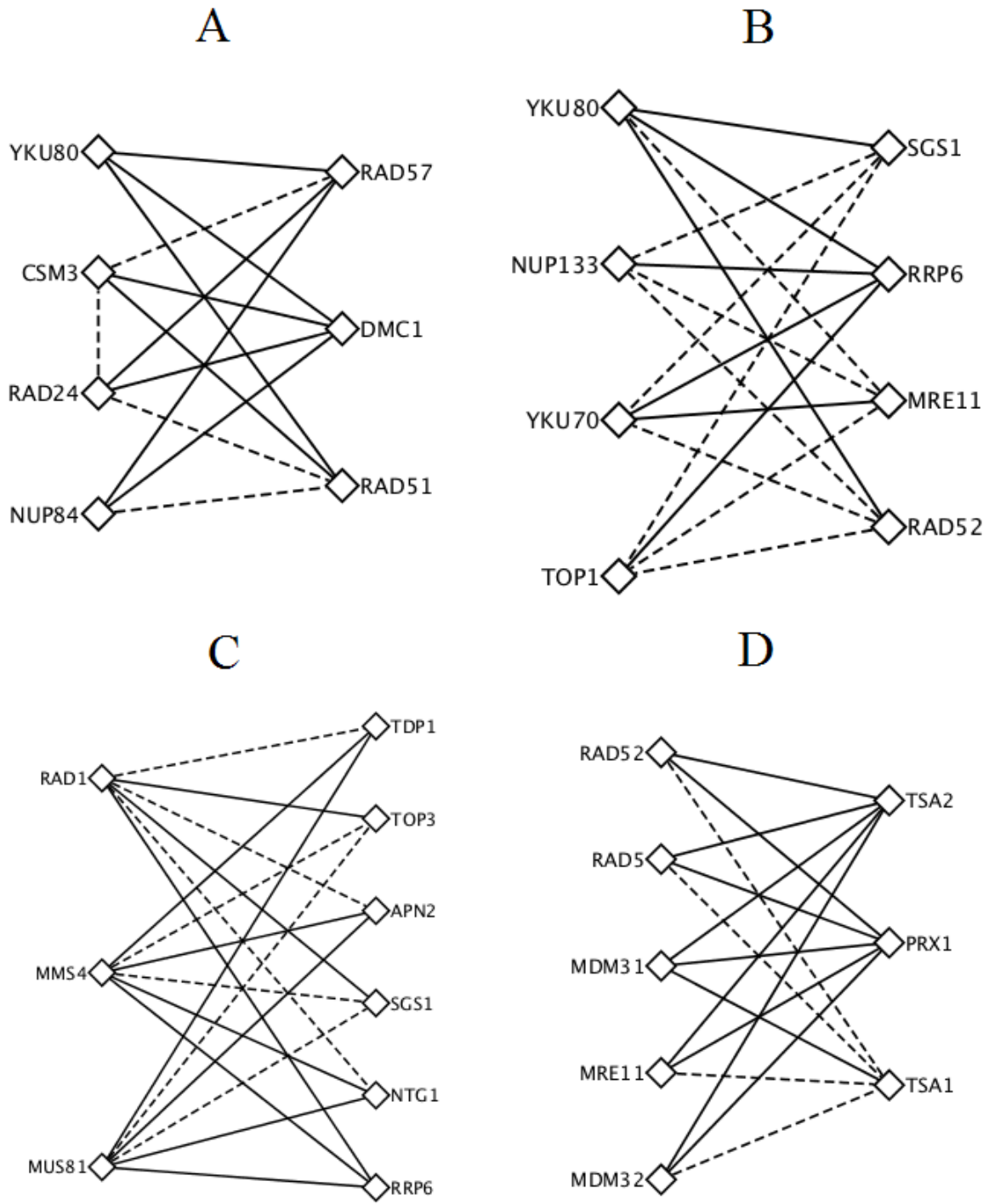

**Table I.** Domain genetic interactions with top ten highest E scores.

| E score | Prob. | Domain$_a$ | Domain$_a$ Pfam ID | # of Yeast proteins with domain$_a$ | Domain$_b$ | Domain$_b$ Pfam ID | # of Yeast proteins with domain$_b$ | # of GI |
|---|---|---|---|---|---|---|---|---|
| 67.48 | 0.46 | PF00225 | Kinesin | 6 | PF01920 | Prefoldin_2 | 4 | 11 |
| 65.59 | 0.72 | PF01302 | CAP_GLY | 4 | PF01920 | Prefoldin_2 | 4 | 12 |
| 63.31 | 0.32 | PF00022 | Actin | 9 | PF01920 | Prefoldin_2 | 4 | 11 |
| 41.73 | 0.025 | PF00071 | Ras | 24 | PF00071 | Ras | 24 | 13 |
| 39.53 | 0.036 | PF00400 | WD40 | 87 | PF01920 | Prefoldin_2 | 4 | 14 |
| 38.41 | 0.12 | PF00071 | Ras | 24 | PF04893 | Yip1 | 3 | 8 |
| 36.53 | 0.027 | PF00069 | Pkinase | 114 | PF01920 | Prefoldin_2 | 4 | 17 |
| 35.15 | 0.35 | PF00225 | Kinesin | 6 | PF02996 | Prefoldin | 3 | 6 |
| 33.16 | 0.80 | PF01920 | Prefoldin_2 | 4 | PF03114 | BAR | 2 | 7 |
| 31.20 | 0.013 | PF00022 | Actin | 9 | PF00069 | Pkinase | 114 | 17 |

**Table II.** Predicted new synthetic lethal genetic interactions with probability greater than 0.9

| Protein 1 | | Protein 2 | | Probability |
|---|---|---|---|---|
| Name | Annotation | Name | Annotation | |
| MYO4 | type V myosin motors | DYN1 | Cytoplasmic heavy chain dynein | 0.9895 |
| RPO21 | RNA polymerase II largest subunit | BRF1 | TFIIIB B-related factor | 0.9816 |
| DST1 | General transcription elongation factor TFIIS | RET1 | Second-largest subunit of RNA polymerase III | 0.9797 |
| STO1 | Large subunit of the nuclear mRNA cap-binding protein complex | RPA190 | RNA polymerase I subunit | 0.9586 |
| STO1 | Large subunit of the nuclear mRNA cap-binding protein complex | RPO21 | RNA polymerase II largest subunit | 0.9586 |
| DST1 | General transcription elongation factor TFIIS | RPA135 | RNA polymerase I subunit A135 | 0.9570 |
| BNR1 | Formin, nucleates the formation of linear actin filaments | DYN1 | Cytoplasmic heavy chain dynein | 0.9441 |
| CEG1 | Alpha (guanylyltransferase) subunit of the mRNA capping enzyme | RPA190 | RNA polymerase I subunit | 0.9437 |
| CEG1 | Alpha (guanylyltransferase) subunit of the mRNA capping enzyme | RPO31 | RNA polymerase III subunit | 0.9437 |
| RPB5 | RNA polymerase subunit | RPA190 | RNA polymerase I subunit | 0.9437 |
| RPB5 | RNA polymerase subunit | RPO31 | RNA polymerase III subunit | 0.9437 |
| DST1 | General transcription elongation factor TFIIS | RPA190 | RNA polymerase I subunit | 0.9265 |
| DST1 | General transcription elongation factor TFIIS | RPO31 | RNA polymerase III subunit | 0.9265 |
| ARO1 | Pentafunctional arom protein | SPT16 | Subunit of the heterodimeric FACT complex | 0.9198 |
| RET1 | Second-largest subunit of RNA polymerase III | SUA7 | Transcription factor TFIIB | 0.9035 |
| BRF1 | TFIIIB B-related factor | RET1 | Second-largest subunit of RNA polymerase III | 0.9035 |
| BRF1 | TFIIIB B-related factor | RPB2 | RNA polymerase II second largest subunit | 0.9035 |

**Table III.** Summary of SLGIs used in this study

| | |
|---|---|
| Number of known SLGIs with no replicates | 10977 |
| Number of unique proteins in known SLGIs | 2640 |
| Number of SLGIs between proteins with domains | 7475 |
| Number of unique proteins in selected SLGIs | 2008 |
| Number of unique domains in selected SLGIs | 1357 |

**Additional files provided with this submission**

There are five supplemental Tables. More Supplement data are available on website www.genenetworks.net.